\documentclass[aps,pre,reprint,superscriptaddress,amsmath,showpacs]{revtex4-1}

\usepackage[breaklinks,colorlinks=true]{hyperref}
\usepackage{graphicx}

\begin{document}

\title{Generalized epidemic process on modular networks}

\author{Kihong Chung}
\affiliation {Department of Physics,
Korea Advanced Institute of Science and Technology, Daejeon
305-701, Korea}

\author{Yongjoo Baek}
\affiliation {Department of Physics,
Korea Advanced Institute of Science and Technology, Daejeon
305-701, Korea}

\author{Daniel Kim}
\affiliation {Department of Physics,
Korea Advanced Institute of Science and Technology, Daejeon
305-701, Korea}

\author{Meesoon Ha}
\email[Corresponding author; ]{msha@chosun.ac.kr}
\affiliation{Department of Physics Education, Chosun University,
Gwangju 501-759, Korea}

\author{Hawoong Jeong}
\affiliation{Department of Physics and Institute for the
BioCentury, Korea Advanced Institute of Science and Technology,
Daejeon 305-701, Korea}
\affiliation{APCTP, Pohang, Gyeongbuk
790-784, Korea}

\date{\today}

\begin{abstract}
Social reinforcement and modular structure are two salient
features observed in the spreading of behavior through social
contacts. In order to investigate the interplay between these two
features, we study the generalized epidemic process on modular
networks with equal-sized finite communities and adjustable
modularity. Using the analytical approach originally applied to
clique-based random networks, we show that the system exhibits a
bond-percolation type continuous phase transition for weak social
reinforcement, whereas a discontinuous phase transition occurs for
sufficiently strong social reinforcement. Our findings are numerically
verified using the finite-size scaling analysis and the crossings of
the bimodality coefficient.
\end{abstract}

\pacs{89.75.Hc, 87.23.Ge, 64.60.aq, 05.70.Fh}


\maketitle

\section{Introduction}
\label{sec:intro}

Numerous studies have discussed epidemics on various types of
networks~\cite{Dorogovtsev2008, Barrat2008, Cohen2010}, but the
epidemics of behaviors in societies remains less understood. The
latter phenomenon has two features that make it more complex than
the former. First, there is {\em social reinforcement}: Each
individual has the memory of previous social interactions, so that
a new behavior is more easily transmitted from an approving
neighbor when it has been affirmed by a greater number of social
connections~\cite{Centola2007, Centola2010, [{There exists a
different perspective on this concept, which considers memoryless
social reinforcement. See }] DeDomenico2013}. Second, human
societies usually have highly-clustered modular
structures~\cite{Granovetter1978}. Developing proper mathematical
devices to deal with these two features has been a theoretical
challenge. In this paper, we present a model for which this
problem can be successfully addressed so that the phase transition
properties of behavioral epidemics can be analytically predicted.

Previous studies have modeled social reinforcement in several
different ways. For example, each node may require a minimum
number of approving neighbors to adopt a
behavior~\cite{Centola2007, Granovetter1978, Watts2002,
Centola2007a}. Alternatively, an individual may go through a
series of intermediate ``awareness'' levels before the final
adoption~\cite{Krapivsky2011, MZheng2013}. Otherwise, the
probability of adoption may change for each recommendation
received from an additional approving neighbor~\cite{LLu2011}. The
{\em generalized epidemic process} (GEP) belongs to the last type,
in which the probability of adoption is an arbitrary function of
the number of previous recommendations~\cite{Janssen2004,
Dodds2004, Bizhani2012}. The GEP on locally tree-like random
networks was analytically solved~\cite{Bizhani2012} using a
self-consistency equation argument analogous to the derivation of
bond percolation threshold on random networks~\cite{Newman2001,
Newman2002}. The solution revealed that the GEP exhibits a
bond-percolation type continuous phase transition for weak social
reinforcement, which changes to a discontinuous transition for
sufficiently strong social reinforcement. Whether the GEP on more
realistic highly clustered networks has similar properties is an
interesting problem, since the previous self-consistency equation
argument assumes locally tree-like structure and cannot be
directly applied to such clustered systems.

While previous works about behavioral epidemics on highly
clustered networks~\cite{Centola2007, Centola2010, Centola2007a,
LLu2011} primarily dealt with the small-world
networks~\cite{Watts1998}, we prefer to focus on the modular
networks consisting of highly-clustered communities joined by
random inter-community links~\cite{Girvan2002}. Such networks
capture three crucial properties of social networks at once:
clustering, modularity, and small worldness. Besides, such modular
networks are structurally very similar to clique-based random
networks, whose cascading properties were recently shown to be
analytically tractable by a variant of self-consistency equation
argument~\cite{Gleeson2009, Hackett2013}. Keeping these
considerations in mind, we have chosen the GEP on modular networks
with equal-sized communities and adjustable proportions of intra-
and inter-community links as the focus of our study. In order to
reduce the complexity of the problem, we simplify the GEP so that
the probability of adoption at and after the second approval is
constant, as was the case in \cite{Janssen2004}.

The rest of the paper is organized into four sections. In
Sec.~\ref{sec:model}, we define the modular network and
the two different kinds (original and modified) of GEP used in
our study. In Sec.~\ref{sec:analysis}, we present an analytical
approach to the original GEP, which predicts outbreak size,
epidemic threshold, and the nature of phase transition. All
predictions can be numerically verified. In Sec.~\ref{sec:results},
with the aid of the crossings of bimodality coefficient,
we numerically show that the modified GEP shows the same
transition nature despite the differences in the model details.
Finally, we summarize our findings and discuss their possible
implications in Sec.~\ref{sec:summary}.

\begin{figure}
\centering
\includegraphics[width=0.9\columnwidth]{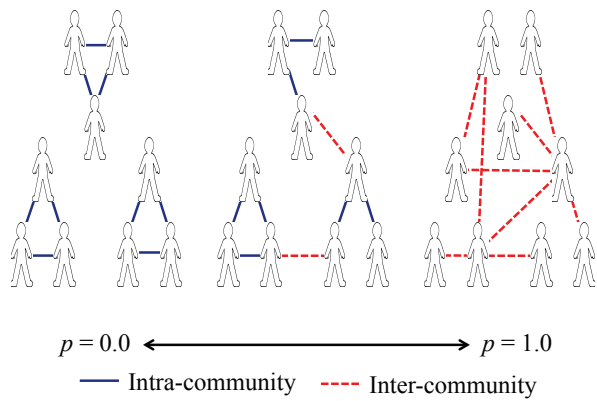}
\caption{\label{fig:mod_net} (Color online) Schematic
illustration of modular networks for various $p$: The network at
$p=0$ consists of $N = 12$ nodes partitioned into communities of
size $c = 3$. As $p$ increases from $0$ to $1$, the disconnected
cliques gradually changes via a series of modular structures to
the ER network.}
\end{figure}

\section{Model}
\label{sec:model}
\subsection{Modular network}

We consider a modular network of $N$ nodes, which consists of
predefined communities with $c$ nodes each. The total number of
links is fixed so that the average degree is exactly $c - 1$. A fraction
$p$ of the links connect random pairs of nodes, while the others are
all intra-community ties. Self-loops and parallel links are forbidden.
If $p = 0$, the network has disconnected cliques of size $c$, whereas
$p = 1$ corresponds to an Erd\H{o}s--R\'{e}nyi (ER) network. Thus, by
varying $p$ between $0$ and $1$, we can adjust the modularity of the
network (see Fig.~\ref{fig:mod_net} for a schematic illustration
of the effect of $p$). This network is very similar to (but not exactly the
same as) the benchmark modular network used in \cite{Girvan2002}. We
define the asymptotic limit of this network as the limit of $N \to \infty$
with $c$ kept finite. In this limit, $p$ is effectively the fraction of
inter-community ties since a random pair of nodes is far more likely to
belong to different communities [$O(N^2)$ pairs] than to the same
community [$O(N)$ pairs].

\subsection{Generalized epidemic process (GEP)}

In the context of behavioral epidemics, the GEP can be defined as
an epidemic process in which the probability of a node adopting
the spreading behavior depends on the number of previous
unsuccessful attempts to persuade the node~\cite{Janssen2004,
Bizhani2012}. In the most general case, the probability of
adoption at the $n$-th attempt can be denoted by $\lambda_n$. In
the beginning, the behavior is adopted by a single random node,
which is {\em active} in the sense that it tries to spread the
behavior to its neighbors. At every instant, one randomly chosen
active node tries to persuade all its neighbors simultaneously,
with the probability of success given by $\lambda_n$; any
persuaded neighbor adopts the behavior and becomes active. After
then, the node {\em deactivates} and no longer participates in the
dynamics, losing interest in spreading the behavior. This process
continues until the system runs out of active nodes. If $R$ nodes
adopt the behavior in the end, we call $R$ the {\em outbreak size}
and $R/N$ the {\em fractional outbreak size}.

In the simplest case when all probabilities of adoption are equal
($\lambda_1 = \ldots = \lambda_\infty = \lambda$),
the process reduces to the well-known susceptible-infected-recovered
(SIR) model~\cite{Newman2002}, which lacks social reinforcement.
We focus on the case minimally more sophisticated than the SIR model,
where $\lambda_1 = \lambda$ and $\lambda_2 = \cdots = \lambda_\infty = T$.
Here $\lambda$ represents the inherent strength of the behavior, while
$T$ indicates the modified strength of the behavior resulting from the
memory of social interactions. When $T > \lambda$, multiple
recommendations synergistically enhance the probability of adoption.
If $T < \lambda$, multiple recommendations have inhibitory effects
on the spreading of behavior. We note that $T < \lambda$ can also be
interpreted as the case when the spreading occurs in an explorative way
rather than an exploitative way, always preferring to spread through
the sites that have been contacted fewer times~\cite{PerezReche2011}.

For a given set of parameters $c$, $p$, and $T$, there exists an
{\em epidemic threshold} $\lambda_c$ such that $R/N$ can converge
to a nonzero value as $N \to \infty$ only for $\lambda > \lambda_c$.
This phenomenon is called the {\em epidemic transition}, which can be
interpreted as an absorbing phase transition whose order parameter is
the largest possible asymptotic limit of $R/N$. Analytic predictions of
this order parameter is explained in Sec.~\ref{sec:analysis}.

\subsection{Modified GEP}

In the original GEP, an active node tries to persuade its neighbors.
In reality, the spreading of behavior may occur in a slightly different
way, in which a randomly chosen susceptible node {\em observes}
its neighbors and decides whether to adopt the behavior depending
on the number of approving neighbors. We require that the node
does not update its state when the number of approving neighbors
has not changed since the last observation. Thus, the process stops
when there is no node that can observe any change in its neighborhood.
This process is different from the original GEP in that the probability
of adoption does not always change sequentially from $\lambda$ to
$T$; depending on the observed number of approving neighbors, the
first probability of adoption experienced by the node may well be $T$
rather than $\lambda$. We call this process the {\em modified} GEP,
whose properties are numerically investigated in Sec.~\ref{sec:results}.

\section{Analysis of original GEP}
\label{sec:analysis}

\begin{figure}
\centering
\includegraphics[width=0.9\columnwidth]{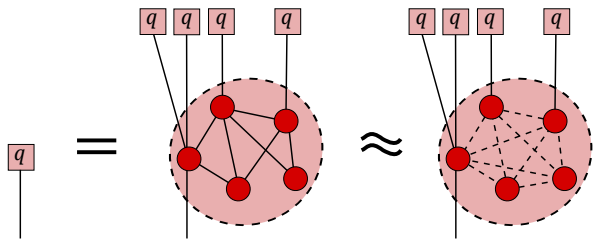}
\caption{\label{fig:sce} (Color online) Schematic illustration of
the self-consistency equation approach to the GEP on modular
networks. A modular network is equivalent to a clique-based
network sharing the same value of $c$ but with every intra-clique
equally disconnected with probability $p$. Since all external
neighbors are similar and independent of each other, a
self-consistency equation can be obtained for the probability
$q$ that an external neighbor adopts the spreading behavior.}
\end{figure}

This section presents a self-consistency equation approach to the
GEP on modular networks, from which the asymptotic limit of $R/N$
can be derived. The approach, originally developed for locally
tree-like networks~\cite{Newman2002,Bizhani2012}, was recently
generalized to clique-based random networks~\cite{Gleeson2009,
Hackett2013}. We describe how this generalized approach can be
adapted to modular networks.

Consider a clique-based random network that consists of cliques
of size $c$ and some additional links that connect random pairs of
nodes belonging to different cliques. The number of these inter-clique
links is such that each node has on average $p(c-1)$
{\em external neighbors} belonging to the other cliques in addition to
$c-1$ {\em internal neighbors} in the same clique. Then, the degree
distribution of the network satisfies
\begin{equation} \label{eq:p}
P(k) = \begin{cases}
\frac{\left[p(c-1)\right]^{k-c+1}}{(k-c+1)!}e^{-p(c-1)} &\text{for $k \ge c-1$}\\
0 &\text{for $k < c - 1$}.
\end{cases}
\end{equation}

We denote by $F_{l,j}$ the probability that a node has adopted the
behavior by the time $l$ internal and $j$ external neighbors have
adopted it. For the sake of simplicity, we assume $T = \lambda$
for the moment. For clique-based random networks, $F_{l,j}$ can be
written as $F_{l+j}$ because both internal and external neighbors
contribute equally to its value~\cite{Hackett2013}. More
specifically, each neighbor fails to persuade the node with
probability $1-\lambda$, thus
\begin{equation}
F_{l+j} = 1 - (1-\lambda)^{l+j}.
\end{equation}
However, in modular networks, on average a fraction $p$ of the
inter-clique links are actually disconnected. To account for these
nonexistent links, the probability that an internal neighbor fails
to persuade the node must be put equal to $p + (1-p)(1-\lambda)$,
where $p$ represents the absence of a link and $(1-p)(1-\lambda)$
is the failure to persuade the node despite the presence of a
link. Therefore,
\begin{equation}
\label{eq:f_lambda}
F_{l,j} = 1 - \left[p + (1-p)(1-\lambda)\right]^l (1-\lambda)^j.
\end{equation}
Returning to the case of $T \neq \lambda$, any $(1-\lambda)^n$
with $n \ge 2$ must be replaced with $(1-\lambda)(1-T)^{n-1}$.
Then, Eq.~(\ref{eq:f_lambda}) is rewritten as
\begin{align}
\label{eq:f}
F_{l,j} = 1 &- [1-(1-p)T]^l (1-T)^{j-1}(1-\lambda) \nonumber \\
&+ \delta_{j,0}\frac{T-\lambda}{1-T}p^l,
\end{align}
which is the expression we use to describe the GEP on modular
networks. We note that the use of Eqs.~(\ref{eq:p}) and (\ref{eq:f})
is the only change required for making the approach of \cite{Gleeson2009,
Hackett2013} applicable to modular networks. The rest of the derivation
is essentially the same as that given by \cite{Hackett2013}, which is
reviewed in the following for the sake of completeness.

We denote by $q$ the probability that a node at an end of a
randomly chosen inter-clique link ends up adopting the behavior
through one of the $k-1$ {\em excess links} other than the
inter-clique link chosen in the beginning. Suppose that the
probability of $j$ among $k-c$ external neighbors ($l$ among $c-1$
internal neighbors) adopting the behavior before the behavior
spreads to the node is given by $B_j^{k-c}(q)$ [$R_l^{c-1}(q)$].
Then $q$ satisfies the self-consistency equation
\begin{equation}
\label{eq:q}
q = \sum_{k=1}^\infty P'(k) \sum_{j=0}^{k-c} \sum_{l=0}^{c-1} B_j^{k-c}(q)
R_l^{c-1}(q) F_{l,j},
\end{equation}
where
\begin{equation}
\label{eq:dist-link}
P'(k) = \frac{(k-c+1)P(k)}{p(c-1)} = P(k-1)
\end{equation}
is the degree distribution of a node at an end of a randomly chosen
inter-clique link (see Fig.~\ref{fig:sce} for a schematic illustration of
this self-consistency equation). Similarly, the probability that a randomly
chosen node adopts the behavior through one of its $k$ links is given by
\begin{equation}
\label{eq:order}
\lim_{N \to \infty} \frac{R}{N} = \sum_{k=0}^\infty P(k) \sum_{j=0}^{k-c+1}
\sum_{l=0}^{c-1} B_j^{k-c+1}(q) R_l^{c-1}(q) F_{l,j}.
\end{equation}
From Eqs.~(\ref{eq:q}) and (\ref{eq:dist-link}), it is clear that
\begin{equation}
\label{eq:order=q}
q = \lim_{N \to \infty} \frac{R}{N}.
\end{equation}
Thus, solving Eq.~(\ref{eq:q}) for $q$ automatically yields the asymptotic
limit of the fractional outbreak size $R/N$.

It is very straightforward to obtain $B_j^{k-c+1}(q)$, since the locally tree-like
structure of inter-clique links implies that each external neighbor adopts the
behavior with independent and identical probability equal to $q$. Thus,
$B_j^{k-c+1}$ is given by the binomial distribution
\begin{equation} \label{eq:binom}
B_j^{k-c}(q) = \binom{k-c}{j}q^j(1-q)^{k-c-j}.
\end{equation}

On the other hand, calculating $R_l^{c-1}(q)$ is far more tedious
due to the strong interdependence arising from local clustering. We
must consider every possible way to spread the behavior among $l$
internal neighbors (while keeping the node itself unaffected) in a single
or multiple steps of cascades. An interested reader is referred to
\cite{Hackett2013} for the detailed derivation, and here we simply describe
the final result:
\begin{equation} \label{eq:r}
R_l^{c-1}(q) = \sum_{(l_i)}\prod_{i = 1}^v B_{l_i}^{n_{c,i}}(\theta_{c,i}).
\end{equation}
The summation is over all possible integer sequences $(l_i)_{i=1}^v$
of variable length $v$ satisfying $l = \sum_{i = 1}^v l_i$ and $l_i > 0$
except that $l_v = 0$ for $l < c - 1$~\cite{[{If we exclude the zero elements,
this set of all $(l_i)$ is just the same as the set of all {\em compositions} of
the integer $l$, which has been extensively studied in combinatorics. See }]
Heubach2009}. Also note that, adopting the simplifying notation $l_0 = 0$
and $m_{i} = \sum_{j = 0}^i l_j$,
\begin{equation}
n_{c, i} = c - 1 - m_{i - 1}
\end{equation}
and
\begin{align} \label{eq:theta}
\theta_{c,i} =
\begin{cases}
\frac{G_{m_{i-1}}^{c-1}(q) - G_{m_{i-2}}^{c-1}(q)}{1 - G_{m_{i-2}}^{c-1}(q)} &\text{if $i \ge 2$}\\
G_0^{c-1}(q) &\text{if $i = 1$},
\end{cases}
\end{align}
where $G_m^{c-1}(q)$ denotes the probability that a single member of the clique
has adopted the behavior by the time $m$ other members have adopted
the behavior:
\begin{align} \label{eq:g}
G_m^{c-1}(q) &= \sum_{k=0}^\infty P(k) \sum_{j = 0}^{k-c+1} B_j^{k-c+1}(q) F_{m,j}
\nonumber \\
&= 1-\frac{1-\lambda}{1-T}[1-(1-p)T]^m e^{-(c-1)pqT} \nonumber \\
&\quad + \frac{T-\lambda}{1-T}p^m e^{-(c-1)pq}.
\end{align}

\begin{figure}[b]
\centering
\includegraphics[width=0.95\columnwidth]{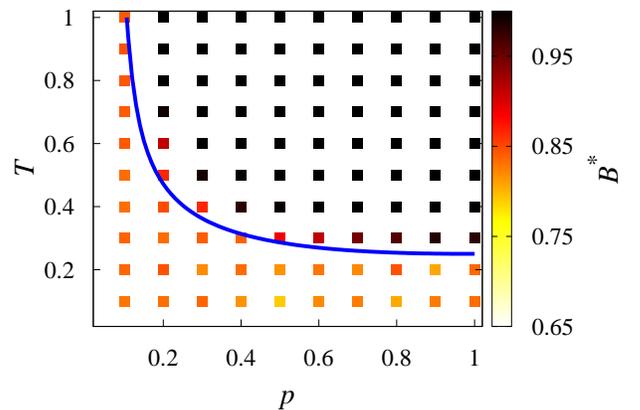}
\caption{\label{fig:diagram} (Color online) Diagram of transition
nature for $c = 6$. The solid line indicates the tricritical line of the original
GEP predicted by the analytical approach, while the gradation of shades
indicates the crossing point values of the bimodality coefficient
$B^*$ for the modified GEP numerically obtained at $N = 768000$
and $1536000$. The curvature of the predicted line is qualitatively
similar to that of the observed boundary.}
\end{figure}

From Eqs.~(\ref{eq:order}), (\ref{eq:order=q}), and (\ref{eq:g}), the
self-consistency equation can be rewritten as
\begin{equation}
\label{eq:q=h}
q = H(q) \equiv \sum_{l=0}^{c-1} R_l^{c-1}(q) G^{c-1}_l(q).
\end{equation}
This equation always has a solution at $q = 0$, which corresponds to
the localized spreading of behavior. An epidemic outbreak is possible
only if Eq.~(\ref{eq:q=h}) has a nonzero solution. In order to check this
possibility, we may examine a Taylor expansion of Eq.~(\ref{eq:q=h})
given by
\begin{align}
\label{eq:poly}
H(q) - q \nonumber &= [H'(0) - 1]q + \frac{H''(0)}{2}q^2 + O(q^3) \nonumber \\
& = a(\lambda - \lambda_c)q + \frac{H''(0)}{2}q^2
+ O[q^3 + q (\lambda - \lambda_c)^2] \nonumber \\
& = 0,
\end{align}
where $H'(0) = 1$ at $\lambda = \lambda_c$ and
$a \equiv \partial H'(0)/\partial \lambda |_{\lambda=\lambda_c}$
is a positive coefficient.

For $H''(0) < 0$, Eq.~(\ref{eq:poly}) implies that $q = 0$ is the only solution
for $\lambda \le \lambda_c$, and a nonzero solution becomes available for
$\lambda > \lambda_c$. When $\lambda$ is only slightly larger than
$\lambda_c$, the nonzero solution satisfies $q \sim (\lambda-\lambda_c)^\beta$
with the critical exponent given by the mean-field value $\beta = 1$. The
other critical exponents cannot be determined from Eq.~(\ref{eq:q=h}) alone,
for they are dependent on the size distribution of finite outbreaks
($R \ll N$)~\cite{Cohen2010}, whose properties are not well described
by Eq.~(\ref{eq:q=h}). Nonetheless, we expect them to be equal to the
mean-field values because, for finite outbreaks, the effect of $T$ is limited
to local communities and would not affect the scale-invariant critical properties.
Thus, whenever $H''(0) < 0$, the epidemic transition would not be
distinguishable from the ordinary bond percolation transition.

By contrast, if $H''(0) > 0$, the transition becomes discontinuous.
The boundary between the two different types of transitions, or
the {\em tricritical point} $(\lambda_t,\,T_t)$, is given by
\begin{equation}
\label{eq:tricritical}
H'(0) = 1, \quad H''(0) = 0.
\end{equation}
For the special case of ER networks ($p = 1$), Eq.~(\ref{eq:tricritical})
yields
\begin{equation}
\lambda_t = 1/(c-1),\quad T_t = \lambda_t/(1-\lambda_t),
\end{equation}
which confirms the result of \cite{Bizhani2012} obtained for
locally tree-like networks. If $p < 1$, the tricritical point
$T_t$ cannot be written in a simple form, but it can still be
calculated by numerically solving Eq.~(\ref{eq:tricritical}).
The predicted relationship between $p$ and $T_t$ is shown
by the thick curve in Fig.~\ref{fig:diagram}. This curve separates
the region of discontinuous transition (above the curve) from that
of continuous transition (below the curve).

\begin{figure}[t]
\centering
\includegraphics[width=\columnwidth]{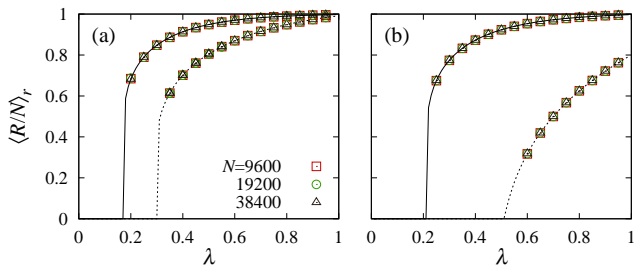}
\caption{\label{fig:size_pred} (Color online) Comparison between
analytical predictions (lines) and numerical estimations (symbols)
of the right-peak average $\langle R/N \rangle_r$, which indicates
the nonzero fractional outbreak size. They give consistent results
for the GEP on both (a) clique-based networks ($c = 6$, $p = 0.2$,
$T = 0.7$ for solid line and $c = 3$, $p = 0.8$, $T = 0.7$ for dashed
line) and (b) modular networks ($c = 6$, $p = 0.2$, $T = 0.7$ for
solid line and $c = 3$, $p = 0.8$, $T = 0.7$ for dashed line).}
\end{figure}

The analytical approach explained in this section can be numerically
verified. The distribution of $R/N$ obtained from simulations consists
of two clearly separable peaks for $\lambda$ sufficiently larger than
$\lambda_c$. Since the left peak is very close to $R/N = 0$, only the
right peak corresponds to the nonzero fractional outbreak size
predicted by the analytical approach. Thus, we may use the right-peak
average of $R/N$, denoted by $\langle R/N \rangle_r$, as an estimator
of the nonzero fractional outbreak size. A comparison between this
estimator (symbols) and the prediction (lines) is shown in
Fig.~\ref{fig:size_pred}. As $N$ increases, the former approaches the
latter for both clique-based [Fig.~\ref{fig:size_pred}(a)] and modular
[Fig.~\ref{fig:size_pred}(b)] networks, which is a strong evidence for
the validity of the theory.

The estimator $\langle R/N \rangle_r$ can be measured reliably only
when the right peak is separable from the left peak and contains
sufficiently many samples. For this reason, the epidemic threshold
$\lambda_c$ cannot be directly verified by $\langle R/N \rangle_r$.
As an alternative indicator of $\lambda_c$, we use the crossing of
bimodality coefficient, which is described in details in
Sec.~\ref{sec:results}. For the moment, we simply note that the
estimated $\lambda_c$ is consistent with the prediction, as shown
for the case of $\lambda = T$ in Fig.~\ref{fig:difference}.

\section{Analysis of modified GEP}
\label{sec:results}

While the analytical approach described in Sec.~\ref{sec:analysis}
has been shown to be accurate for the ordinary GEP, it is inaccurate
for the modified GEP. Since multiple persuasion attempts in the former
process can be treated as a single observation event in the latter, the
modified GEP always has a fewer number of updates. Thus, its outbreak
size must be smaller than that of the ordinary GEP, and consequently its
epidemic threshold would be larger. However, we may still check whether
these differences lead to any notable changes in the nature of phase
transition, namely its (dis)continuity and critical exponents. For this
purpose, we perform numerical simulations of the modified GEP on
modular networks. Provided that $p$ is high enough to ensure the
existence of a percolating giant cluster, the qualitative properties of
the model were checked to be largely invariant with respect to the
choice of $c$ and $p$. Therefore, without loss of generality, we present
the results for $c = 6$ and $p = 0.2$ unless otherwise noted.

\begin{figure}[t]
\centering
\includegraphics[width=0.95\columnwidth]{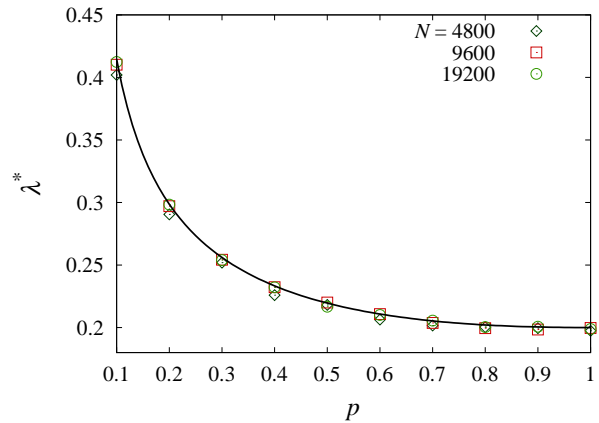}
\caption{\label{fig:difference} (Color online) Comparison of $\lambda^*$ (symbols)
obtained from the crossings of $B(\lambda)$ with the analytical prediction of
$\lambda_c$ (line) in the absence of social reinforcement ($T = \lambda$) for
the GEP on modular networks with $c = 6$. The prediction is
in good agreement with the numerical results.}
\end{figure}

\begin{figure}
\centering
\includegraphics[width=0.95\columnwidth]{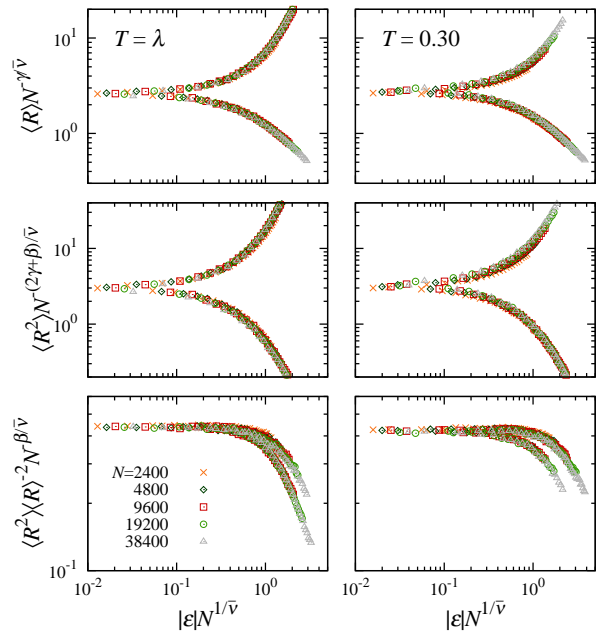}
\caption{\label{fig:fss} (Color online) FSS of moments of outbreak
size $R$ for the modified GEP with $c = 6$ and $p = 0.2$. The notation
$\epsilon \equiv \lambda/\lambda_c - 1$ is used. The data are obtained
from $10^5$ samples. When social reinforcement is absent ($T = \lambda$,
$\lambda_c = 0.3283$; left panel) or weak ($T = 0.30$, $\lambda_c
= 0.3384$; right panel), the data are well collapsed with $\beta =
1$, $\gamma = 1$, and $\bar{\nu} = 3$ as expected.}
\end{figure}

\begin{figure*}
\centering
\includegraphics[width=0.95\textwidth]{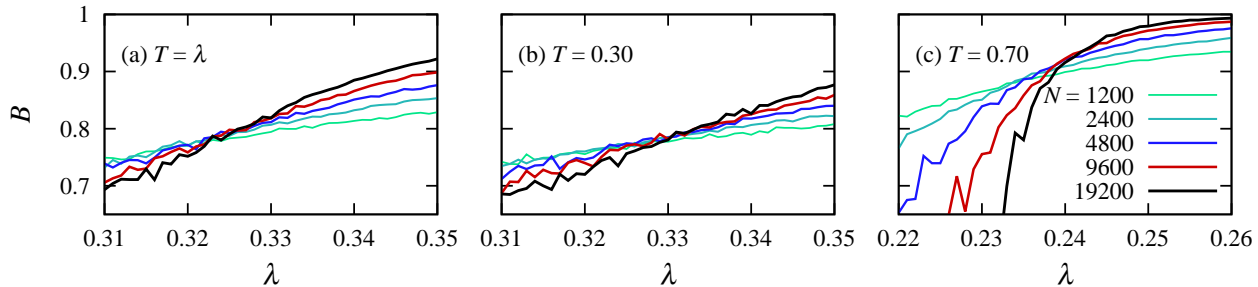}
\caption{\label{fig:lambda_bimod} (Color online) Bimodality
coefficient $B(\lambda)$ of the outbreak size distribution near
$\lambda_c$ for the modified GEP with $c = 6$ and $p = 0.2$,
measured for $10^{5}$ samples. The slope of the curve increases
with $N$, making a series of crossings that stay around a value
much smaller than $1$ for (a) the absence of social reinforcement
($T = \lambda$) and (b) weak social reinforcement ($T = 0.30$),
while it is close to $1$ for (c) strong social reinforcement ($T = 0.70$).}
\end{figure*}

\begin{figure}
\centering
\includegraphics[width=0.9\columnwidth]{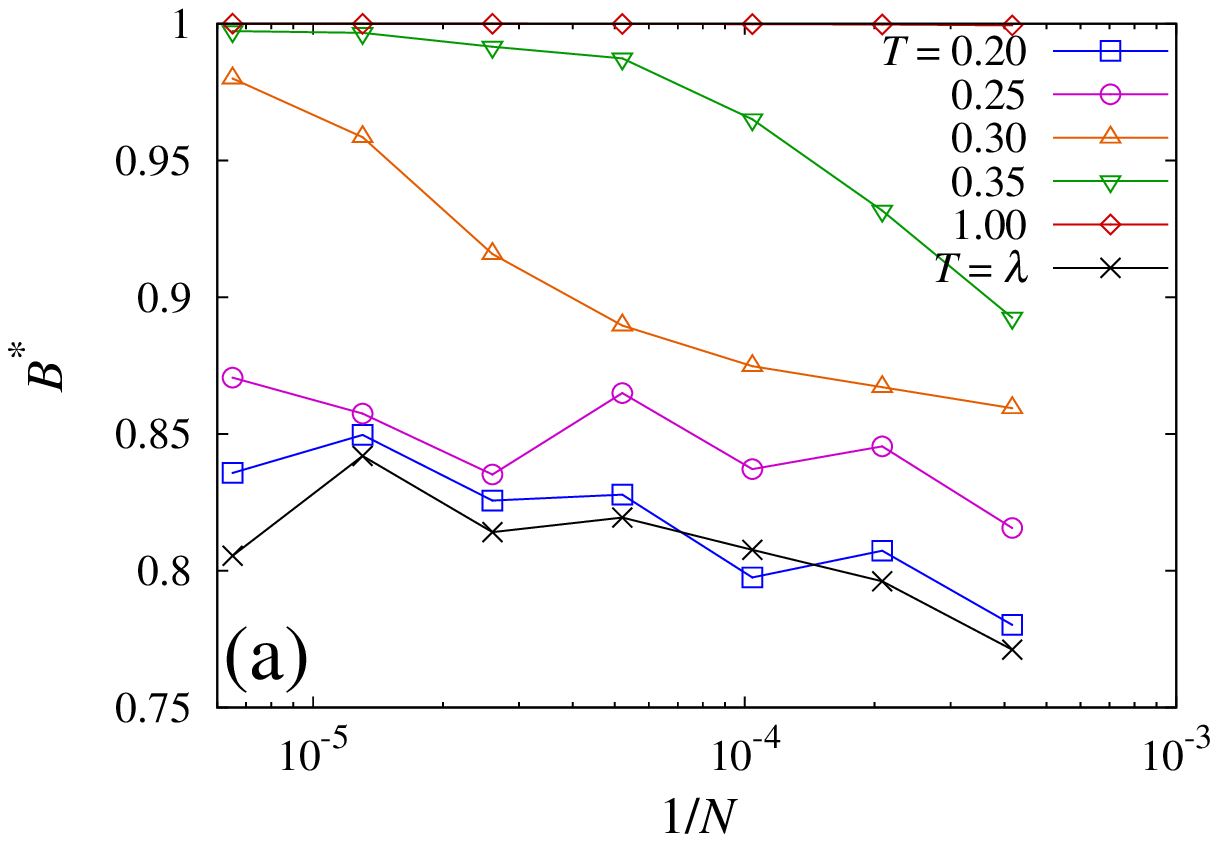}
\includegraphics[width=0.9\columnwidth]{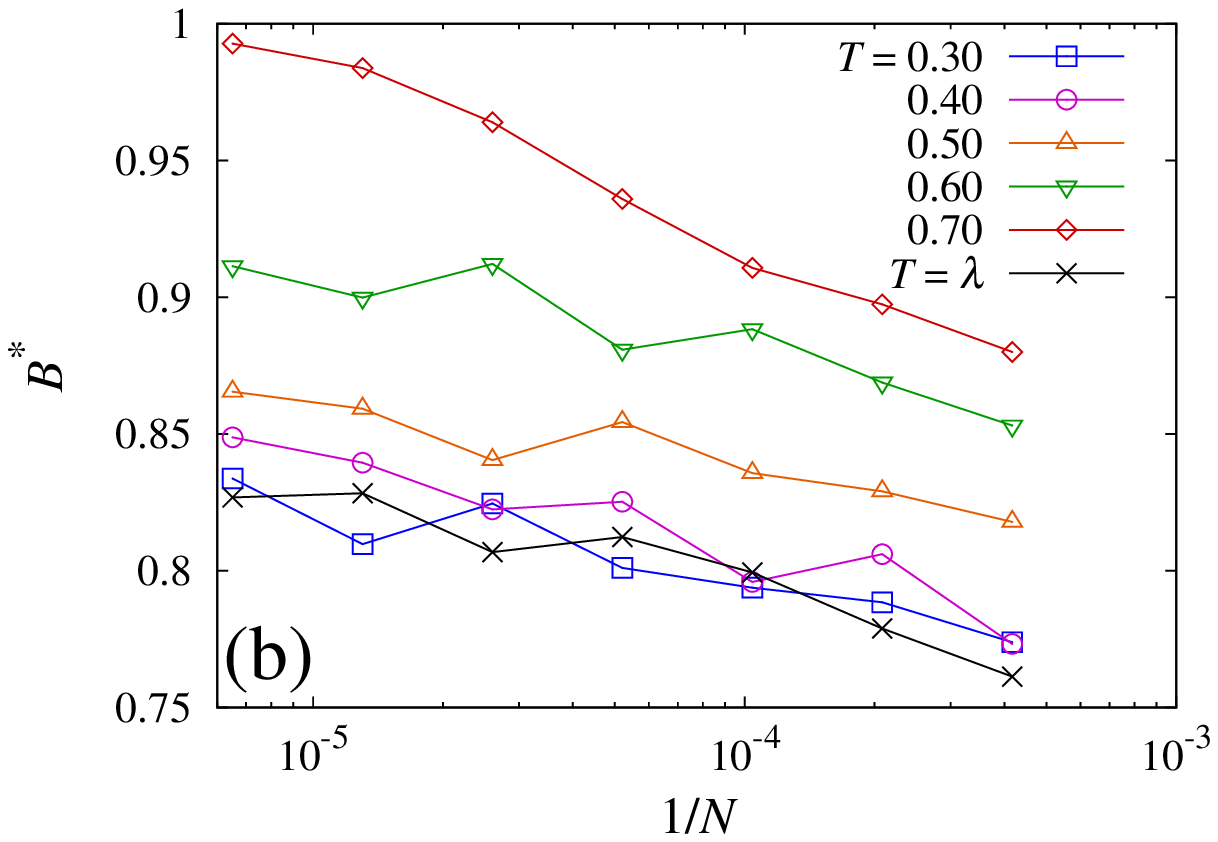}
\caption{\label{fig:n_bimod} (Color online) The rightmost
crossings $B^*(N)$ of two $B(\lambda)$ curves for system sizes $N$
and $2N$, obtained from $10^5$ runs of the modified GEP with $c = 6$.
Results for (a) $p = 1.0$ and (b) $p = 0.2$ are shown separately. Interpreting
the asymptotic limit of $B^*$ as $B(\lambda_c)$, we observe that
$B(\lambda_c)$ approaches $1$ only for sufficiently strong social
reinforcement.}
\end{figure}

We first estimate the critical exponents by the FSS analysis of the
moments of outbreak size $R$. Using the notation $\epsilon \equiv
(\lambda - \lambda_c) / \lambda_c$, Fig.~\ref{fig:fss} shows that
all data are well collapsed by the predicted FSS
forms~\cite{DeSouza2011}
\begin{align}
\langle R \rangle &= N^{\gamma/\bar{\nu}} f(\left|\epsilon\right|N^{1/\bar{\nu}}), \\
\langle R^2 \rangle &= N^{(\beta+2\gamma)/\bar{\nu}} g(\left|\epsilon\right|N^{1/\bar{\nu}})
\end{align}
provided that social reinforcement is either absent ($T =
\lambda$) or sufficiently weak ($T = 0.30$) to yield continuous
transitions. Here $\lambda_c$ is chosen to give the best collapse,
and critical exponents are those of the mean-field bond
percolation universality class ($\beta = 1$, $\gamma = 1$,
$\bar{\nu} = 3$). This shows that the critical properties of the modified
GEP are likely to be the same as those of the original GEP.

Now we move on to check whether the modified GEP has both continuous
and discontinuous transitions. In order to numerically determine the
transition nature, we employ the bimodality coefficient $B$~\cite{SAS12.3}
of the outbreak size distribution, which is defined as
\begin{equation}
B(\lambda) = \frac{\left\langle \left( \frac{R - \langle R
\rangle}{\sigma} \right)^3 \right\rangle^2 + 1}{\left\langle
\left( \frac{R - \langle R \rangle}{\sigma} \right)^4
\right\rangle},
\end{equation}
where $\sigma \equiv \sqrt{\langle R^2 \rangle - \langle R \rangle^2}$.
This coefficient is equal to $1$ only for random variables with
exactly two possible choices, such as Bernoulli trials or two
delta peaks. It is smaller than $1$ for the other kinds of
distributions~\cite{Pearson1921}. In the case of the modified GEP,
as $N \to \infty$, the bimodality coefficient at the transition
point $B(\lambda_c)$ converges to $1$ only for discontinuous
transitions, whereas it approaches a smaller value determined by
the outbreak size distribution $P(R) \sim R^{-\tau+1}$ (which is equivalent
to the probability of a randomly chosen node belonging to a cluster of size
$R$ in the original percolation problem~\cite{DeSouza2011}) for continuous
transitions:
\begin{equation} \label{eq:b_cont}
B(\lambda_c) = \frac{(4-\tau)(6-\tau)}{(5-\tau)^2}.
\end{equation}
For the mean-field bond-percolation universality class, we have $\tau = 5/2$,
so $B(\lambda_c) = 0.84$.

As shown in Fig.~\ref{fig:lambda_bimod}, a pair of $B(\lambda)$ curves
measured at different values of $N$ make some crossings. To estimate
$\lambda_c$ and $B(\lambda_c)$, we keep track of one of those crossings
(say the rightmost one, to remove the ambiguity) as $N$ is successively
increased by factors of $2$. Denoting by $(\lambda^*,\,B^*)$ the coordinates
of the rightmost crossing, Figs.~\ref{fig:difference} and \ref{fig:n_bimod} show
their behavior as $N$ is successively increased by factors of $2$. It seems
that both $\lambda^*$ and $B^*$ converge to a limit, which can be interpreted
as $\lambda_c$ and $B(\lambda_c)$, respectively. The agreement between
the predicted $\lambda_c$ and the estimator $\lambda^*$ for the original
GEP shown in Fig.~\ref{fig:difference} supports this interpretation.

As shown for both $p = 1$ [Fig.~\ref{fig:n_bimod}(a)] and $p = 0.2$
[Fig.~\ref{fig:n_bimod}(b)], $B^*$ converges to values near $0.84$
when $T$ is sufficiently small or fixed equal to $\lambda$, which is
to be expected from Eq.~(\ref{eq:b_cont}) and the mean-field bond-percolation
critical exponents implied by Fig.~\ref{fig:fss}. Meanwhile,
for sufficiently large values of $T$, $B^*$ approaches very close to $1$,
suggesting discontinuous transitions. Although the curves obtained at
intermediate values of $T$ seemingly converge to values other than $0.84$
and $1$, we expect this to be a crossover phenomenon at finite size.

Based on these observations, we use $B^*$ obtained from two $B(\lambda)$
curves with sufficiently large values of $N$ as a proxy for the true
$B(\lambda_c)$. Figure~\ref{fig:diagram} shows the $(p,T)$-dependence of
$B^*$ obtained from $N = 768000$ and $N = 1536000$ by a gradation of
shades. Both the dark region of discontinuous transition and the light region
of continuous transition are observed, whose border shows a curvature
similar to that of the tricritical line (solid curve) of the original GEP. Thus,
despite the differences between the original and modified GEP, many properties
related to the nature of phase transition remain similar.

\section{Summary and discussions}
\label{sec:summary}

Using the self-consistency equation method adapted to clique-based
random networks, we formulated an analytical approach to the
generalized epidemic process (GEP) on modular networks, which
describes the emergence of collective behavior through social
contacts. The theory predicts that strong social reinforcement
induces an abrupt epidemic outbreak (discontinuous transition),
while weak social reinforcement leads to a gradual epidemic
outbreak (continuous transition) whose critical exponents are
those of the mean-field bond percolation universality class. We
also obtained the mathematical condition
[Eq.~(\ref{eq:tricritical})] for the tricritical line that forms
the boundary between the two different types of transition. These
analytical predictions were numerically confirmed by tracking the
crossings of the bimodality coefficient test. We also numerically
showed that the modified GEP, in which the nodes observe rather
than persuade others, exhibits the same transition nature despite
the differences in the model details.

We note that the modularity parameter $p$ changes the positions of
tricritical points only, leaving the nature of critical points and
the existence of discontinuous transitions unaffected. Thus, the
modular networks ($p < 1$) turn out to have the same critical
properties as those of the Erd\H{o}s--R\'{e}nyi (ER) networks ($p
= 1$). It is also notable that the tricritical point $T_t$
increases monotonically as $p$ is decreased. Since lowering $p$
increases clustering, it could have amplified the effect of social
reinforcement so that $T_t$ is minimized at some $p < 1$. Instead,
$T_t$ is minimized at $p = 1$, for which the clustering
coefficient becomes zero. All these properties might be
attributable to the fact that modular networks look similar to ER
random networks when coarse grained, since clustering is localized
to small communities and long-range inter-community links are
randomly distributed. Reducing $p$ (i.e., removing inter-community
links) is just equivalent to lowering the average degree of the
coarse-grained ER random network, which merely raises the critical
and the tricritical points. The critical exponents, which
represent the scale-invariant properties~\cite{Goldenfeld1992},
would remain unchanged as long as the coarse-grained structure of
the network is essentially the same.

Finally, we comment on a few possible extensions of our study.
First, we only considered the case when degree and community
size distributions are very homogeneous, while both distributions
tend to be heterogeneous for realistic networks~\cite{Guimera2003}.
Just as bond percolation on scale-free networks has a unique set of
critical exponents, the critical properties of the GEP would be affected
by such structural heterogeneity. Second, we discussed the GEP only
from the phase transition perspective, but the model can also be studied
from the viewpoint of optimization, as was done for small-world
networks~\cite{LLu2011}. For example, what is the optimal value
of $p$ that maximizes the outbreak size for given $\lambda$ and
$T$? The analytical approach of Sec.~\ref{sec:analysis} can also
be applied to answer such questions. Lastly, the tricritical line predicted
in our study might have its own critical properties, which can be
addressed in future studies.

\begin{acknowledgments}
This research was supported by Basic Science Research Program
through the National Research Foundation of Korea(NRF) funded
by the Ministry of Science, ICT and Future Planning(No.
2011-0011550)(M.H.); (No. 2011-0028908)(K.C., Y.B., D.K., H.J.).
\end{acknowledgments}

\bibliography{EL11225}

\end{document}